\documentclass[aps,twocolumn,prl,reprint]{revtex4}
 
\usepackage{amsmath,amssymb} 
\usepackage{bm} %
 
\usepackage{color} %
\usepackage{graphicx}

\usepackage[colorlinks]{hyperref} %
\hypersetup{
citecolor=blue,
linkcolor=blue,
urlcolor=blue
}

\usepackage{epstopdf}
\newcommand{\fig}[1]{Fig.~\ref{#1}}

\newcommand{\eq}[1]{Eq.~(\ref{#1})}

\begin{document}
\title{
A 750 GeV dark matter messenger   at  the Galactic Center
}
\author{
Xian-Jun Huang,\ %
Wei-Hong Zhang,\ %
Yu-Feng Zhou\\  %
\textit{Key Laboratory of Theoretical Physics,}\\
\textit{Institute of  Theoretical Physics,
Chinese Academy of Sciences}\\
\textit{Beijing, 100190, P.R. China}\\
}
\date{\today}
\begin{abstract}
The first data from the LHC Run-2  have shown 
a possible excess in diphoton events with 
invariant mass $\sim 750$~GeV,
suggesting the existence of  a new resonance 
which may decay dominantly into dark matter (DM) particles. 
We show in a simple model that 
the reported diphoton excess at the LHC is  consistent with 
another photon excess, 
the  $2$~GeV excess in cosmic gamma-ray fluxes towards 
the Galactic Center  observed by  the Fermi-LAT.
Both the excesses can be simultaneously explained by 
a $\sim 60$~GeV scalar DM particle 
annihilating dominantly into two gluons with 
a typical thermal  annihilation cross section,
which leads to the prediction of
a  width to mass ratio $\Gamma/M\approx \mathcal{O}(10^{-2})$ of 
the resonance. 
The upper limit on the dijet search at LHC Run-1 leads to 
a $lower$ limit on the predicted cross section for 
DM annihilating  into $\gamma\gamma$ final states
$\langle\sigma v\rangle_{\gamma\gamma} \gtrsim\mathcal{O}(10^{-30})~\mbox{cm}^{3}\mbox{s}^{-1}$. 
Both the predictions can be
tested  by the LHC, Fermi-LAT and future experiments.
\end{abstract}
\maketitle %
\newpage
{\bf Introduction.}
Recently, 
the ATLAS and CMS collaborations have reported 
the first data  from the LHC Run-2 at  $\sqrt{s}=13$~TeV,
based on the integrated luminosity of 
3.2~fb$^{-1}$ and 2.6~fb$^{-1}$, respectively
\cite{LHC-diphoton}. 
Both the collaborations have shown 
a possible excess in diphoton events, %
suggesting the existence of a new resonance particle
$\phi$ with mass  $M\approx 750$~GeV.
The distribution of the observed events at ATLAS  favours
a width to mass ratio of the resonance $\Gamma/M\approx 0.06$ with
a local (global) significance of 3.9~$\sigma$ (2.6~$\sigma$).
The CMS collaboration has reported 
a mild peak at $\sim760$~GeV with 
a local (global) significance of $2.6\sigma$ ($1.2\sigma$)
and  slightly favours  a narrow width. 
Assuming  a large width, 
the ATLAS (CMS) data favour a production cross section
$10\pm 3~\mbox{fb}~(6\pm3~\mbox{fb})$
\cite{
Franceschini:2015kwy%
}.
Other analyses assuming narrow width give
$\sim 6.2 \ (5.6)$~fb 
for ATLAS (CMS)
\cite{
Buttazzo:2015txu%
}.
Recent updates from both ATLAS and CMS have shown that
a mild upward fluctuation at 750 GeV also exists in the Run-1 data at 8~TeV.
The local (global) significance of the diphoton excess in 
the combined CMS data of 8+13~TeV has increased to 
3.4~$\sigma$ (1.6~$\sigma$) 
\cite{LHC-diphoton-CMS-1603}.

The excess,  
if  not due to statistic fluctuations,
can be an intriguing  clue of new physics beyond the standard model (SM):
the resonance $\phi$ should not be the only new particle.
If the observed number of diphoton events are explained by 
the usual loop processes involving $only$ the SM particles,
$\phi$ should decay into these SM particles appearing in the loop with large rates,
which  is inconsistent with the LHC Run-1 data
\cite{
Low:2015qep,
Agrawal:2015dbf%
}.
Furthermore, 
$\phi$ should have extra  tree-level invisible decays if the large width reported by ATLAS
is confirmed.
An interesting  possibility is  that 
the dark matter (DM) particle is among the decay final states of $\phi$
\cite{
Mambrini:2015wyu,
Backovic:2015fnp,
Franceschini:2015kwy,%
Bi:2015uqd,
Barducci:2015gtd,
Bai:2015nbs,
Han:2015cty,
Bauer:2015boy,
Dev:2015isx,
Davoudiasl:2015cuo,%
D'Eramo:2016mgv%
}.
In this scenario, 
$\phi$ plays a role of messenger connecting 
the invisible and visible sectors by
making the DM particles couple to 
gluons and photons through $\phi$-exchange, 
which has rich phenomenological consequences.

Note that there is another photon related excess.
Recently,
a number of groups have  independently
found statistically strong evidence of an excess in cosmic gamma-ray fluxes at $\sim2$~GeV towards
the inner regions  around the Galactic center (GC) from the data of Fermi-LAT
\cite{Goodenough:2009gk,
Hooper:2010mq,
Boyarsky:2010dr,
Abazajian:2010zy,
Hooper:2011ti,
Abazajian:2012pn,
Gordon:2013vta,
Macias:2013vya,
Abazajian:2014fta,%
Hooper:2013rwa,
Huang:2013pda,
Daylan:2014rsa,
Calore:2014xka,%
TheFermi-LAT:2015kwa%
}.
The morphology  of this GC excess (GCE) emission is consistent with 
a spherical emission profile expected from DM annihilation
\cite{%
Abazajian:2012pn,
Gordon:2013vta,
Hooper:2013rwa,
Calore:2014xka%
}.
The determined energy spectrum of the excess emission
is in general compatible with  
a DM %
particle self annihilating into 
$b \bar b $ final states with 
a  
typical thermal annihilation cross section 
\cite{
Hooper:2013rwa,
Daylan:2014rsa%
}.
Plausible astrophysical explanations also exist,
such as the unresolved point sources of mili-second pulsars
\cite{
Abazajian:2010zy,%
Hooper:2011ti,%
Abazajian:2012pn,%
Gordon:2013vta,%
Bartels:2015aea,%
Lee:2015fea%
}
and
the interactions between the cosmic rays and the molecular gas
\cite{Macias:2013vya,%
Abazajian:2014fta,%
Gaggero:2015nsa%
}.

In this work,
we show that the two reported photon excesses can be closely connected.
They can be simultaneously explained by
a simple scalar DM model with a light DM particle mass  $\sim60$~GeV and 
a typical thermal annihilation cross section, which leads to the predictions that
$i$) the resonance should have a large width, $\Gamma/M\gtrsim\mathcal{O}(10^{-2})$ from 
the required  DM mass and annihilation cross section;
$ii$) %
the upper limit on the dijet search at LHC Run-1 leads to 
a $lower$ limit on the predicted  cross section $\langle\sigma v\rangle_{\gamma\gamma} \gtrsim\mathcal{O}(10^{-30})~\mbox{cm}^{3}\mbox{s}^{-1}$ for DM annihilating into $\gamma\gamma$ with a line-shape gamma-ray spectrum.
Both of them can be tested by the LHC,  Fermi-LAT and future experiments.

{\bf Effective interactions.}
We consider a simple model where 
the resonance $\phi$ is a pseudo-scalar particle and 
the DM particle $\chi$ with mass $m_{\chi}$ is a real scalar.
The interactions related to $\phi$ and $\chi$ is given by
\begin{align}
\mathcal{L}_{\phi\chi} \supset &
\frac{1}{2}(\partial_{\mu} \phi)^{2}
+\frac{1}{2}(\partial_{\mu} \chi)^{2}
-\frac{1}{2}M^{2}\phi^{2}
\nonumber\\
&-\frac{1}{2}m_{\chi}^{2}\chi^{2}
-\frac{1}{2} g_{\chi}\phi\chi^{2},
\end{align}
where $g_{\chi}$  is the dimensionful $\phi\chi\chi$coupling strength. 
The  resonance  $\phi$  can couple to the SM gauge fields typically through
loop processes (see e.g. Refs.~\cite{
others%
}).
Since $\phi$ is much heavier than the electroweak (EW) scale, 
we start with effective dimension-five EW gauge-invariant interactions 
\begin{align}\label{eq:EWinteraction}
\mathcal{L}
\supset
\frac{g^{2}_{1}}{2\Lambda}\phi B_{\mu\nu}\tilde{B}^{\mu\nu}
+\frac{g^{2}_{2}}{2\Lambda}\phi W_{\mu\nu}^{}\tilde{W}^{\mu\nu}
+\frac{g^{2}_{g}}{2\Lambda}\phi G_{\mu\nu}^{}\tilde{G}^{\mu\nu},
\end{align}
where for the gauge fields 
$\tilde{F}_{\mu\nu}=\frac{1}{2}\epsilon_{\mu\nu\alpha\beta}F^{\alpha\beta}$,
$g_{1,2,g}$ are the dimensionless effective coupling strengths, and 
$\Lambda$ is a common energy scale. 
After the EW symmetry breaking,
the interaction terms involving physical EW gauge bosons $A$, $Z$ and $W^{\pm}$ are
\begin{align}
\mathcal{L}
\supset &
\frac{g^{2}_{A}}{2\Lambda}\phi A_{\mu\nu}\tilde{A}^{\mu\nu}
+\frac{g^{2}_{Z}}{2\Lambda}\phi Z_{\mu\nu}\tilde{Z}^{\mu\nu}
+\frac{g^{2}_{AZ}}{2\Lambda}\phi A_{\mu\nu}\tilde{Z}^{\mu\nu}
\nonumber\\
&+\frac{g^{2}_{W}}{2\Lambda}\phi W_{\mu\nu}\tilde{W}^{\mu\nu}
+\frac{g^{2}_{g}}{2\Lambda}\phi G_{\mu\nu}\tilde{G}^{\mu\nu},
\end{align}
where the couplings $g_{A,Z,ZA,W}$ are related to the couplings in \eq{eq:EWinteraction} as
$g^{2}_{A}=g^{2}_{1}c_{W}^{2}+g^{2}_{2}s_{W}^{2}$, 
$g^{2}_{Z}=g^{2}_{1}s_{W}^{2}+g^{2}_{2}c_{W}^{2}$, 
$g^{2}_{ZA}=2s_{W}c_{W}(g^{2}_{2}-g^{2}_{1})$, and 
$g^{2}_{W}=g^{2}_{2}$
with $s_{W}^{2}=1-c_{W}^2=\sin^{2}\theta_{W}\approx 0.23$.
The partial decay widths for the decays $\phi\to\gamma\gamma$, $gg$ and 
$\chi\chi$ are given by
\begin{align}\label{eq:widths}
\frac{\Gamma_{\gamma\gamma}}{M}
&=
\pi\alpha_{A}^{2}\left(\frac{M}{\Lambda}\right)^{2}, \
\frac{\Gamma_{gg}}{M}=8\pi\alpha^{2}_{g}\left(\frac{M}{\Lambda}\right)^{2},
\nonumber\\
\frac{\Gamma_{\chi\chi}}{M}&=\frac{g_{\chi}^{2}\beta_{\chi}}{32\pi M^{2}} ,
\end{align}
respectively, 
where $\alpha_{A,g}=g^{2}_{A,g}/4\pi$, and 
$\beta_{\chi}=(1-4m_{\chi}^{2}/M^{2})^{1/2}$ is 
the velocity of the final state DM particles 
in the $\phi$ rest frame.

The UV origins of the  pseudoscalar $\phi$ can be
axion-like particles from the breaking of the Peccei-Quinn symmetry
\cite{PQsym},
pesudo-Goldstone boson from composite Higgs models
\cite{
Gripaios:2009pe%
},
or
from the extended Higgs sectors such as the two-Higgs-doublet models.
If $\phi$ is a SM singlet and couples to the SM gauge bosons through 
new vector-like heavy fermions which have 
small mixings with the SM fermions,
the constraints from the oblique parameters $S$ and $T$, 
the EW precision test,
and the flavor physics can be evaded
\cite{
Ellis:2015oso%
}.
Since $\phi$ is a pseudo-scalar, 
it does not mix directly with the SM Higgs boson.
Thus is less constrained by the measured properties of the SM Higgs boson.
Furthermore,
the DM-nucleus scattering matrix element for gluons
$\langle N |G_{\mu\nu}^{a}\tilde{G}^{a\mu\nu}| N\rangle$ is  vanishing  as 
the operator $G_{\mu\nu}^{a}\tilde{G}^{a\mu\nu}$ is CP-odd, 
which makes the DM particles easily evade 
the stringent  constraints  from DM direct detection experiments.

{\bf Diphoton excess.}
We shall focus on the case 
where the reported $\gamma\gamma$ excess
at the LHC is generated  by 
gluon-fusion through the  $s$-channel $\phi$-exchange.
Other non-resonant mechanisms have also been considered 
( see e.g. in Refs.~\cite{nonRes}).
In the narrow width approximation,
the production cross section for diphoton (dijet) is given by 
\begin{align}\label{eq:diphoton}
\sigma_{\gamma\gamma(jj)} \approx\frac{C_{gg}}{s (\Gamma/M) }
\left(\frac{\Gamma_{gg}}{M}\right)
\left(\frac{\Gamma_{\gamma\gamma(gg)}}{M}\right) ,
\end{align}
where the coefficient $C_{gg}$ incorporates  the convolution  over 
the gluon parton distribution functions of the proton.
At  $\sqrt{s}=13\ (8)$ TeV, 
$C_{gg}\approx2137\ (174)$ 
\cite{
Franceschini:2015kwy%
}. 
Higher order QCD corrections can be taken into account  by 
the $K$-factors with typically $K_{gg}\approx1.48$.
Making use of  \eq{eq:widths},
the products of  the couplings required to reproduce 
the diphoton excess can be estimated as
\begin{align}
\left(\frac{\alpha_{A}}{0.01}\right)^{2}
\left(\frac{\alpha_{g}}{0.1}\right)^{2}
\left(\frac{M/\Lambda}{0.18}\right)^{4}
\approx %
\left(\frac{\sigma_{\gamma\gamma}}{8\ \text{fb}}\right)
\left(\frac{\Gamma/M}{0.06}\right) .
\end{align}
Thus %
the common scale $\Lambda$ can still be larger  than  the mass of the resonance $\phi$,
although not significantly larger. 
For weakly coupled models, 
large effective couplings can be obtained %
by  introducing multiple heavy intermediate particles running in the loop.
A large total width $\Gamma/M\sim 0.06$ can be obtained by 
additional $\phi$ decay channels, such as decay into DM particles.
Including the invisible decay $\phi\to \chi\chi$, the total width is 
\begin{align}\label{eq:totalwidth}
\Gamma=\Gamma_{\chi\chi}+\Gamma_{gg}+\kappa \Gamma_{\gamma\gamma} ,
\end{align}
where $\kappa=1+(g_{Z}^{4}+g_{ZA}^{4}/2+2g_{W}^{4})/g_{A}^{4}$.
If the total width $\Gamma$ is dominated by $\Gamma_{\chi\chi}$,
a large $\Gamma/M\approx 0.06$ requires an effective coupling
$g_{\chi}^{2}/(4\pi M^{2})\approx 0.5$ which is  large but   
still within the perturbative regime.

\begin{figure*}
\begin{center}
\includegraphics[width=0.3\textwidth]{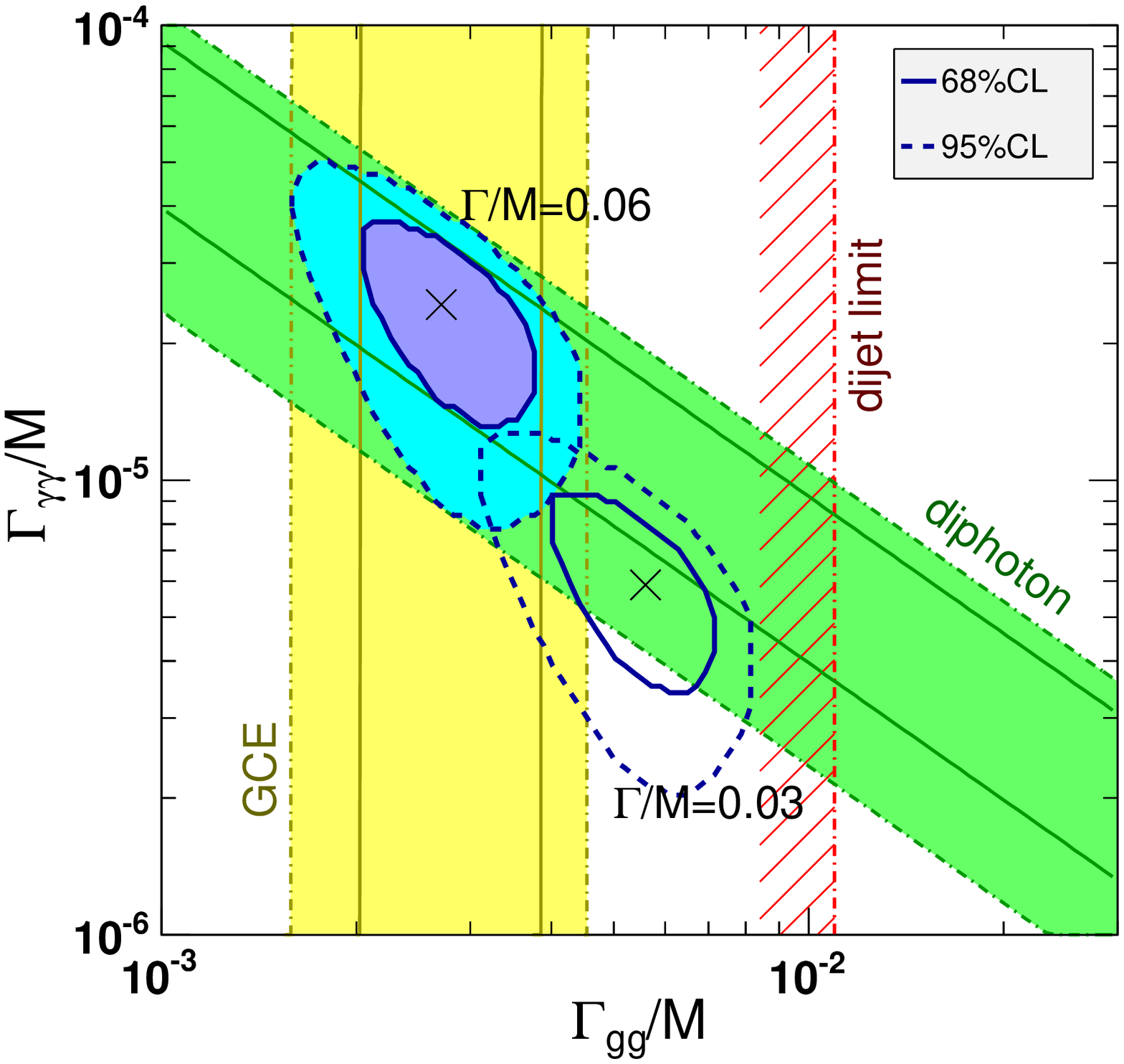}
\includegraphics[width=0.3\textwidth]{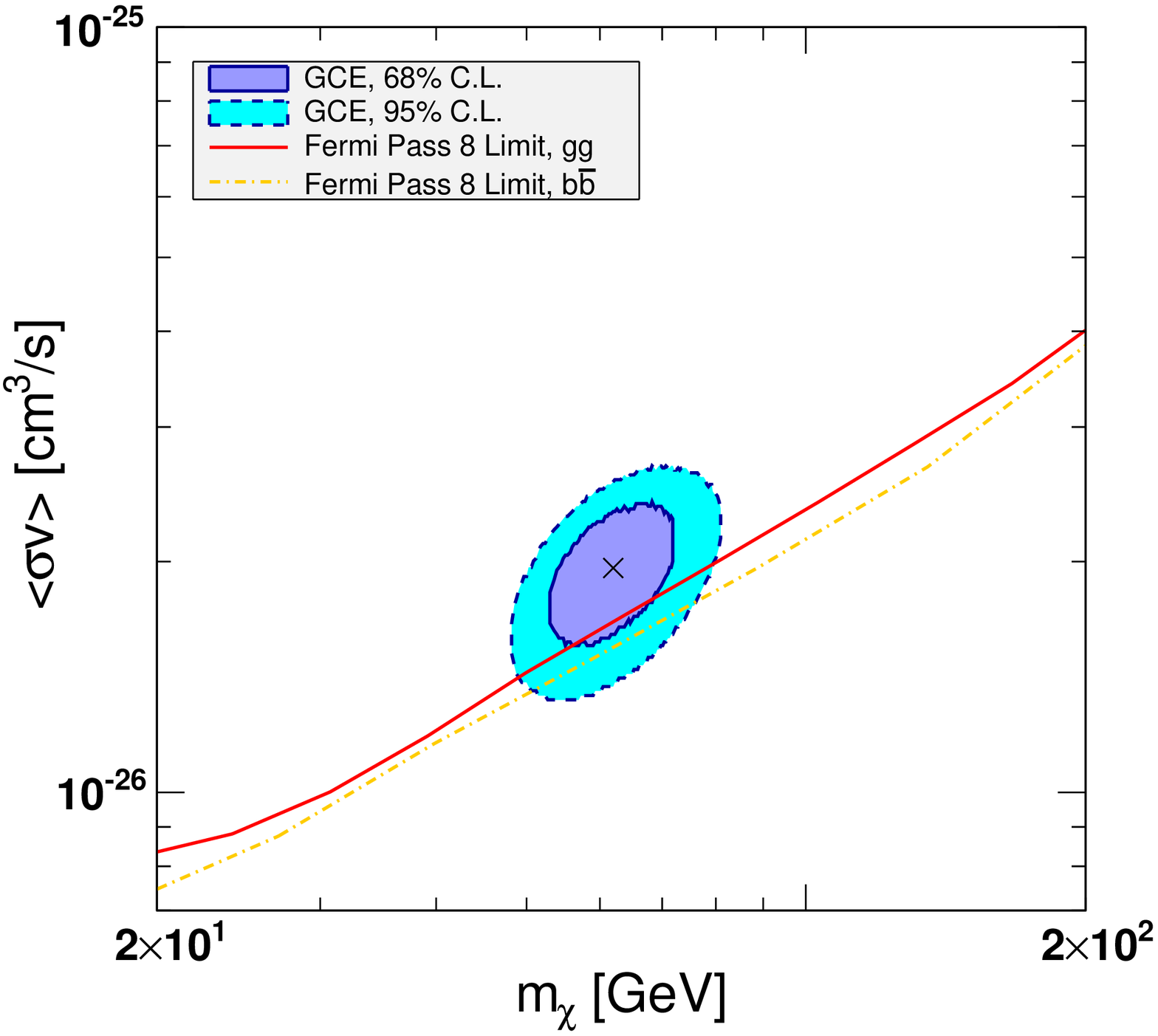}
\includegraphics[width=0.3\textwidth]{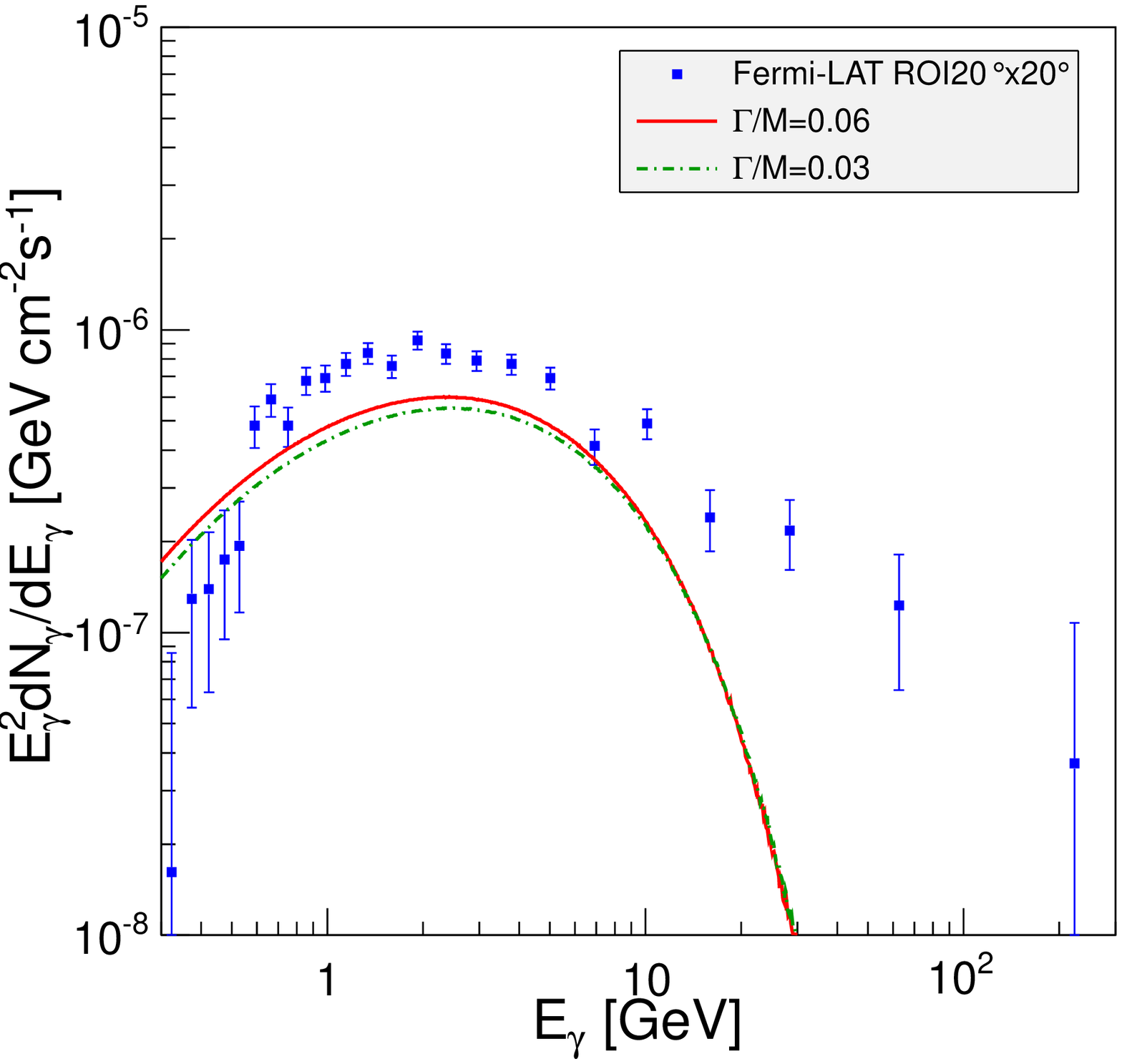}
\caption{
Left)
Allowed regions in $(\Gamma_{gg}/M$, $\Gamma_{\gamma\gamma}/M)$ plane at 
$68\%$ C.L. (violet contour) and $95\%$ C.L. (light-blue contour)  from
the combined fit to the data of diphoton excess
\cite{LHC-diphoton}, 
GCE~\cite{Calore:2014xka} 
and 
the constraints from the Run-1 dijet search limits 
\cite{Aad:2014aqa},
for $\Gamma/M=0.06$.  
The regions  allowed by the individual experiment  at $95\%$~C.L. are also shown.
The open contours correspond to a similar fit with $\Gamma/M=0.03$.
Middle) 
Allowed regions for the parameters 
$(m_{\chi}, \langle \sigma v\rangle_{gg})$ from the fit with $\Gamma/M=0.06$, 
together with the conservative upper  limits at $95\%$~C.L. (solid curve) derived from 
the  Fermi-LAT data on  the gamma rays of dSphs for $b\bar b$ channel (dashed curve)
\cite{
Ackermann:2015zua%
}.
See text for details.
Right)
Energy spectra of the gamma-ray fluxes from the best-fit parameters in \eq{eq:bestfit}
for $\Gamma/M=0.06$ and 0.03, respectively, 
together with the GCE data derived in 
Ref.~\cite{
Calore:2014xka%
}. 
}
\label{fig:GCE}
\end{center}
\end{figure*}

Constraints already arise from the LHC Run-1 data on the searches for 
general resonances. For instance, 
$\sigma_{ZZ}\lesssim 12\mbox{ fb}$
\cite{
Aad:2015kna%
},
$\sigma_{Z\gamma}\lesssim 4.0\mbox{ fb}$
\cite{
Aad:2014fha%
},
and 
$\sigma_{WW}\lesssim 40\mbox{ fb}$
\cite{
Khachatryan:2015cwa,%
Aad:2015agg%
}.
For the LHC Run-2  with $\phi$ produced from gluon fusion,
it is expected that these upper bounds will be relaxed roughly by 
a factor 
$r=0.38 C_{gg}(13\mbox{ TeV})/C_{gg}(8\mbox{ TeV})\approx 4.7$. 
The  coupling between $\phi$ and the gluons is directly constrained by 
the null results of the search for 
dijet from a generic resonance at the Run-1,  
$\sigma_{jj} \lesssim 2.5$~pb%
~\cite{Aad:2014aqa}.
If $\Gamma$ is dominated by $gg$ final states, 
one obtains a stringent limit $\Gamma_{gg}/M \lesssim 1.6\times 10^{-3}$.
However, if $\Gamma$ is dominated by $\Gamma_{\chi\chi}$, 
the constraint can be significantly weaker.
For $\Gamma/M\approx0.06$, we find
$\Gamma_{gg}/M \lesssim 1.1\times 10^{-2}$.
In this work,
we consider a representative  case where 
$\phi$ couples dominantly to the $U(1)$ gauge field, i.e. $g_{2}=0$.
In this case, the $ZZ(Z\gamma)$ channel is suppressed as
$\sigma_{ZZ(Z\gamma)}/\sigma_{\gamma\gamma}=0.09\ (0.6)$.
Note that in the opposite case where $g_{1}=0$, 
the cross section $\sigma_{ZZ(Z\gamma)}$
is enhanced  by a factor of 11~(6.7), 
which is already severely constrained by the Run-1 data.

{\bf GC excess.}
In this model, 
the velocity-averaged DM annihilation  cross section multiplied by 
the DM relative velocity for two gluons (photons) final states is given by
\begin{align}\label{eq:sigmav}
\langle \sigma v \rangle_{gg(\gamma\gamma)} 
\approx 
\frac{256\pi m_{\chi}^{2}
\left(\frac{\Gamma_{\chi\chi}}{M} \right)
\left(\frac{\Gamma_{gg(\gamma\gamma)}}{M} \right)
}{ \left[(M^{2}-4m_{\chi}^{2})^{2}+M^{2}\Gamma^{2}\right] \beta_{\chi}
} ,
\end{align}
where %
we have neglected  the $p$-wave contributions.
We shall perform a combined  $\chi^{2}$-analysis  to both 
the diphoton excess at the LHC and 
the GCE from the Fermi-LAT  to see if they can be consistently
explained by a common parameter set 
$\{\Gamma_{gg}/M$, $\Gamma_{\chi\chi}/M$, $m_{\chi}\}$, 
for fixed values of $\Gamma/M$.
For the diphoton excess,
we take a naively  weighted average of ATLAS and CMS results 
$\sigma_{\gamma\gamma}=8\pm 2.1$~fb.
The upper limit from the dijet process is taken into account  by 
constructing a $\chi^{2}$-term %
corresponding to the $95\%$ upper limit  
at  Run-1,
assuming Gaussian distribution.
The GCE data are taken from  Ref.
\cite{
Calore:2014xka%
}.
The gamma-ray fluxes are calculated and averaged over
a square region of interest (ROI) $20^{\circ}\times 20^{\circ}$ in the sky with 
latitude $|b|<2^{\circ}$ masked out.
In the calculation,
we adopt a contracted NFW profile with inner slop $\gamma=1.26$,
as suggested by the observed morphology of the gamma-ray emission
\cite{%
Abazajian:2012pn,
Gordon:2013vta,
Hooper:2013rwa,
Calore:2014xka%
},
and is  
normalized to the local DM density $\rho_{0}=0.4\mbox{ GeV}\cdot\mbox{cm}^{-3}$. 
The halo DM annihilation into $gg$ generates  diffuse gamma rays with 
a broad energy spectrum due to hadronization.
The injection spectrum %
for DM annihilating into two gluons
are generated by Pythia 8.201~\cite{Sjostrand:2007gs}.

We first perform a fit to the data of GCE alone. 
The result shows that a  cross section close to the typical thermal cross section is favoured
$\langle \sigma v \rangle_{gg} = (1.96^{+0.26}_{-0.24})\times 10^{-26}~\mbox{cm}^{3}\mbox{s}^{-1}$
with a  relatively small DM particle mass $m_{\chi}=62.0^{+6.6}_{-6.3}~\mbox{GeV}$.
The goodness of fit $\chi^{2}/\text{d.o.f}=24.6/22$ indicates 
a good agreement with the data for DM annihilation into two gluons.
A consequence of the required  DM mass and annihilation cross section is that 
the total width $\Gamma$ of $\phi$ cannot be too small.
From the definition of $\Gamma$, 
it follows   that 
$\Gamma^{2} \gtrsim 4\Gamma_{gg}\Gamma_{\chi\chi}$,
and a lower bound on the total width can be derived  from \eq{eq:sigmav}
\begin{align}\label{eq:widthlimit}
\frac{\Gamma}{M} \gtrsim 0.023
\left(\frac{60~\mbox{GeV}}{m_{\chi}}\right)
\left(\frac{M}{750~\mbox{GeV}}\right)^{2}
\frac{\langle \sigma v \rangle_{gg}^{1/2}}{\langle \sigma v \rangle_{0}^{1/2}} ,
\end{align}
where $\langle \sigma v \rangle_{0}=1.5\times 10^{-26}~\mbox{cm}^{3}\mbox{s}^{-1}$ is
the cross section  at the $2\sigma$ lower bound allowed by the GCE data.
Thus a consistent explanation for the diphoton excess and the CGE predicts 
a minimal required value of $\Gamma/M \approx \mathcal{O}(10^{-2})$ 
which is favoured by the current ATLAS data, 
and can be tested by CMS and  the future data.

In the combined fit, 
we consider  two choices of total width, 
$\Gamma/M=0.06$  favoured by the ATLAS data, and 
$\Gamma/M=0.03$ which is close to the minimal allowed value by the GCE data.
The results of the determined parameters are as follows
\begin{align}\label{eq:bestfit}
\Gamma_{gg}/M&=2.7\pm0.4 ~(5.6^{+0.9}_{-1.0})\times 10^{-3} ,
\nonumber\\
\Gamma_{\gamma\gamma}/M&=2.4^{+0.8}_{-0.7}~(0.59^{+0.20}_{-0.17})\times 10^{-5} ,
\\
m_{\chi}&=63.7^{+6.6}_{-6.3}~(65.9^{+6.3}_{-4.9})~\mbox{GeV} ,
\nonumber
\end{align}
with $\chi^{2}/\mbox{d.o.f}=24.7/24$ $(26.9/24)$ for  $\Gamma=0.06$ $(0.03)$.
The allowed regions in 
the $(\Gamma_{gg}/M,\Gamma_{\gamma\gamma}/M)$ plane at  
$68\%$ and $95\%$ C.L. for two parameters, 
corresponding to $\Delta \chi^{2}=2.3$ and 6.0, respectively, 
are shown in the left panel of \fig{fig:GCE},
together with the allowed regions by each individual experiment.
For $\Gamma/M=0.06$, 
the total $\chi^{2}$ is almost unchanged compared with 
that from the fit to the GCE data alone,  
which shows that 
the diphoton excess, GCE and dijet limits can be made consistent 
with each other within this this model. 
While for $\Gamma/M=0.03$, 
a slightly larger $\chi^{2}$ is obtained, 
which is mainly due to the tension between the  Run-1 dijet constraint 
and the total width as can be seen from \eq{eq:widthlimit}.

In the middle panel of  \fig{fig:GCE} we show
the allowed regions for the parameters 
$(m_{\chi}, \langle \sigma v\rangle_{gg})$ at $68\%$ and $95\%$ C.L..
At present, 
the most stringent constraint on the DM annihilation cross sections
are provide by the Fermi-LAT data on the diffuse gamma rays
of the dwarf spheroidal satellite galaxies (dSphs)
\cite{
Ackermann:2015zua%
}.
We make a conservative estimation of the limit on 
the cross section for the $gg$ final states
based on the known limit on that for the $b\bar b$ final states from  the 6-year Fermi-LAT data
as follows.
For a gamma-ray  spectrum $dN^{(b\bar b)}_{\gamma}/dE$ generated from 
DM annihilation into $b\bar b$ with given values of 
$m_{\chi}$ and $\langle \sigma v\rangle_{b\bar b}$,
we search for a  cross section $\langle \sigma v\rangle_{gg}$ for the $gg$ channel  with 
a DM particle mass $m_{\chi}'$ 
which satisfies the condition that 
$dN^{gg}_{\gamma}/dE$ is just above  $dN^{(b\bar b)}_{\gamma}/dE$ for $all$ the gamma-ray energies.
The Fermi-LAT limit on $\langle \sigma v\rangle_{b\bar b}$ at $m_{\chi}$
is then estimated  as a  conservative limit on $\langle \sigma v\rangle_{gg}$ at 
$m_{\chi}'$.
The resulting limits are shown in \fig{fig:GCE}, together with that for the $b\bar b $ final states.
As can be seen from the figure, the two limits are quit similar.  
There is a possible tension between the GCE favoured regions and the Fermi-LAT limits.
Note that 
in the analysis of the Fermi-LAT collaboration, 
the uncertainties in the $J$-factors were taken into account assuming 
a  NFW type parametrization of the DM density profile. 
A recent analysis directly using the spherical Jeans equations rather than taking a
parametric DM density profile as input showed that
the $J$-factor can be smaller by a factor about $2-4$ for 
the case of Ursa Minor, 
which relaxes the constraints on the DM annihilation cross section to
the same amount
\cite{
Ullio:2016kvy%
}.
In the right panel of \fig{fig:GCE}, we show
the gamma-ray spectra for the best-fit parameters in \eq{eq:bestfit}, 
together with the Fermi-LAT data
\cite{
Calore:2014xka%
}.
Although the predicted spectra look slightly  lower than 
the data, %
the obtained  $\chi^{2}$ values do  indicate
good agreements with the data. This is because
the correlations between the data points not shown in the figure have been considered 
\cite{
Calore:2014nla%
}.

The annihilation of halo DM also generates extra cosmic-ray antiparticles 
such as antiprotons and positrons. 
Compared with  prompt gamma-rays, 
the prediction for cosmic-ray charged particles  from DM annihilation suffers from 
large uncertainties in the  cosmic-ray propagation models.
For a DM particle mass $\sim 60$~GeV, 
the predicted $\bar p/p$ ratio peaks at a lower energy $\sim10$~GeV, 
which suffers from additional uncertainties due to the effect of solar modulation.
The upper limits on the DM annihilation cross section 
from the AMS-02 and PAMELA data on $\bar p/p$ ratio for various channels 
have been studied  for a number of propagation models 
and DM density profiles ( see e.g. 
\cite{
Jin:2015sqa,%
Jin:2015mka,%
Hooper:2014ysa,
Jin:2012jn%
}). 
In general, the obtained limits are weaker than that derived from 
the gamma rays of dSphs.  
Only in the extreme case with the ``MAX'' propagation model 
\cite{
Jin:2014ica,%
Jin:2013nta%
} 
where the propagation parameters are  adjusted to  
generate maximal  antiproton flux 
while still be consistent with other comic-ray observables 
such as the B/C flux ratio,
the upper limits  from $\bar p/p$ can be compatible with that from the gamma rays 
for DM particle mass below $\sim 100$ GeV.
The constraints from the cosmic-ray positrons are strongly dependent on 
the annihilation final states. 
For leptonic final states such as $e^{+}e^{-}$ and $\mu^{+}\mu^{-}$, 
the derived upper limits from the AMS-02 positron flux can reach the typical thermal cross section 
for DM particle mass below 50--100~GeV
\cite{ 
Ibarra:2013zia%
}.
But for hadronic final states such as $b\bar b$, the corresponding limits are rather weak,
typically at $\mathcal{O}(10^{-24})~\text{cm}^{3}\text{s}^{-1}$.
The $gg$ final state generates a  softer positron spectrum in comparison with 
the $b\bar b$ final states. 
Thus the corresponding limits are expected to be even weaker.

\begin{figure}[thb]
\begin{center}
\includegraphics[width=0.7\columnwidth]{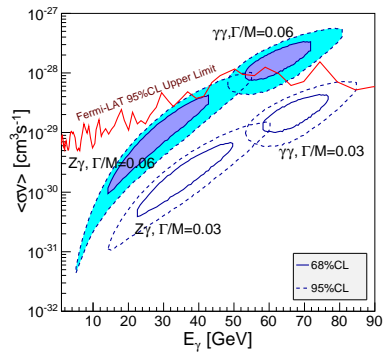}
\caption{
Predictions for $\langle \sigma v \rangle_{\gamma\gamma}$ and 
$\langle \sigma v \rangle_{Z\gamma}$ as a function of photon energy
using the allowed  parameters from the fit to the data of diphoton exces and GCE with 
dijiet limits included, for $\Gamma/M=0.06$ and 0.03, respectively.
The exclusion limits of Fermi-LAT 
\cite{
Ackermann:2015lka%
} for region R16 
are also shown.
}
\label{fig:gamma-line}
\end{center}
\end{figure}
 
{\bf Gamma-ray lines.}
Since $\phi$ couples to two photons, 
the DM particles inevitably annihilate into 
$\gamma\gamma$ ($Z \gamma$ if $g_{1}\neq g_{2}$) with 
line-shape energy spectra at 
$m_{\chi}$  ($m_{\chi}(1-m_{Z}^{2}/4m_{\chi}^{2})$ for $Z\gamma$),
which is difficult to be mimicked  by  conventional astrophysical contributions.
The diphoton produced  at LHC and from halo DM annihilation are strongly correlated.
From \eq{eq:diphoton} and (\ref{eq:sigmav}) it follows that
$\sigma_{\gamma\gamma}/\sigma_{jj}= \langle \sigma v \rangle_{\gamma\gamma}/\langle \sigma v \rangle_{gg}$. 
Therefore,  a $lower$ limit on 
$\langle \sigma v \rangle_{\gamma\gamma}$ can be derived from 
the upper limit on the dijet production cross section
\begin{eqnarray}\label{eq:lowerlimit}
\langle \sigma v \rangle_{\gamma\gamma}
&\gtrsim& ~4.8\times 10^{-30}~\mbox{cm}^{3}\mbox{s}^{-1} 
\left(\frac{\sigma_{\gamma\gamma}}{3.8~\mbox{fb}}\right)
 \nonumber \\
&&
~\times
\left(\frac{12~\mbox{pb}}{\sigma_{jj}}\right) 
\left(\frac{r}{4.7}\right)
\frac{\langle \sigma v \rangle_{gg}}{\langle \sigma v \rangle_{0}} ,
\end{eqnarray}
where the reference value for $\sigma_{\gamma\gamma}$ is at its $2\sigma$ lower bound.
The limit is roughly an order of magnitude lower than
the current Fermi-LAT sensitivity. %
Making use of  the determined parameters 
we obtain the predictions for $\langle \sigma v \rangle_{\gamma\gamma,Z\gamma}$
for $\Gamma/M=0.06$ and 0.03 as shown in 
\fig{fig:gamma-line}.
The cross section for $Z\gamma$ channel is related to that of $\gamma\gamma$ as
$\langle \sigma v \rangle_{Z\gamma} \approx
\langle \sigma v \rangle_{\gamma\gamma}
(1-m_{Z}^{2}/4m_{\chi}^{2})^{3}g_{ZA}^{2}/2g_{A}^{2}$.
For a comparison, the current  $95\%$~C.L. limits  from 
the Fermi-LAT gamma line search based on  Pass-8 data
\cite{
Ackermann:2015lka%
}
for the ROI $16^{\circ}\times 16^{\circ}$ (R16) is also shown in \fig{fig:gamma-line}.
Since the Fermi-LAT limits are obtained assuming the Einasto profile, 
they are rescaled by  a factor of $0.52$ to compensate the differences in 
the  J-factors, as we have adopted a contracted NFW profile.
For $\Gamma/M=0.06~(0.03)$, 
the predicted typical cross section is 
$\mathcal{O}(10^{-28})~(\mathcal{O}(10^{-29}))~\mbox{cm}^{3}\mbox{s}^{-1}$.
As can be seen from \fig{fig:gamma-line},
a significant portion of the parameter space is constrained  by 
the current Fermi-LAT data for $\Gamma/M=0.06$. But there is ample parameter space
for lower values of  $\Gamma/M$, such as for $\Gamma/M=0.03$.
Future  experiments such as CALET and DAMPE are able to reach  the lower limit 
of the cross section derived in \eq{eq:lowerlimit}
in the near future with larger statistics and higher energy resolutions.

In summary, 
we have shown that the reported photon excesses at LHC and GC can be 
simultaneously explained by a simple DM model. 
The best-fit  DM particle mass is around 60~GeV and 
the annihilation cross section is typically thermal, 
which predicts that $\Gamma/M$ of the resonance $\phi$ should be
at least of $\mathcal{O}(10^{-2})$.
The model  predicts 
a minimal  cross section of $\mathcal{O}(10^{-30})~\mbox{cm}^{3}\mbox{s}^{-1}$
for DM annihilating into $\gamma\gamma $ which results in  line-shape spectrum. 
Both of them are testable by the LHC, Fermi-LAT and future DM indirect detection experiments.

{\bf Acknowledgements.}
This work is supported 
by
the NSFC  under Grants
No. 11335012 and
No. 11475237.

\bibliographystyle{apsrev} %
\bibliography{diphotonGC4,misc}
\end{document}